# Stress-induced Eukaryotic Translational Regulatory Mechanisms

Dilawar Ahmad Mir[1], Zhengxin Ma[1], Jordan Horrocks[1], Aric N Rogers[1]

[1] Kathryn W. Davis Center for Regenerative Biology and Aging, Mount Desert Island Biological Laboratory, Bar Harbor, ME

* Corresponding Authors: dilawerjk@gmail.com

MDI Biological Laboratory, 159 Old Bar Harbor Rd, Bar Harbor, ME 04609

Phone: 1-207-288-9880 ext 436

*Abstract*

The eukaryotic protein synthesis process entails intricate stages governed by diverse mechanisms to tightly regulate translation. Translational regulation during stress is pivotal for maintaining cellular homeostasis, ensuring the accurate expression of essential proteins crucial for survival. This selective translational control mechanism is integral to cellular adaptation and resilience under adverse conditions. This review manuscript explores various mechanisms involved in selective translational regulation, focusing on mRNA-specific and global regulatory processes. Key aspects of translational control include translation initiation, which is often a rate-limiting step, and involves the formation of the eIF4F complex and recruitment of mRNA to ribosomes. Regulation of translation initiation factors, such as eIF4E, eIF4E2, and eIF2, through phosphorylation and interactions with binding proteins, modulates translation efficiency under stress conditions. This review also highlights the control of translation initiation through factors like the eIF4F complex and the ternary complex and also underscores the importance of eIF2α phosphorylation in stress granule formation and cellular stress responses. Additionally, the impact of amino acid deprivation, mTOR signaling, and ribosome biogenesis on translation regulation and cellular adaptation to stress is also discussed. Understanding the intricate mechanisms of translational regulation during stress provides insights into cellular adaptation mechanisms and potential therapeutic targets for various diseases, offering valuable avenues for addressing conditions associated with dysregulated protein synthesis.

## *Introduction*

The stress response in cells triggers a cascade of changes in gene regulation, encompassing transcription, mRNA processing, and translation [1-3]. Among these changes, translation stands out as the primary mechanism affected by cellular stress [4]. Adapting translation becomes imperative to confront the challenges imposed by stressful cellular conditions. The regulatory components of the translational machinery play a pivotal role in governing protein synthesis. Particularly, initiation emerges as a crucial target for regulation during stress, given its significant influence on the overall translation rate. Even in smaller organisms, the translation initiation mechanism involves over a hundred macromolecules. Maintaining the utmost accuracy in protein synthesis is essential for cellular functions, necessitating the highest level of translational fidelity.

The first and most direct mechanism impacted by cellular stress is translation [5]. Translation adaptation is necessary in order to react to the challenges and obstacles imposed by conditions of cellular stress. In the regulation of protein synthesis, translational machinery components are actively involved. Initiation is the most important target of regulation under stress, consistent with its crucial role in determining the global rate of translation [6]. Translational machinery components are actively participating in the regulation of proteins synthesis. Even in small microorganisms, the translation initiation mechanism requires more than one hundred macromolecules. Accuracy of protein synthesis is vital for life; consequently, to achieve the requirement of cellular functions, the highest degree of fidelity of translation is mandatory. Various mechanisms elicited within cells lead to a global repression of translation under diverse stress conditions, primarily occurring during the initiation process [3,7-9].

Repression of cap recognition and the reprocessing of the ternary complex through eIF2α phosphorylation are critical steps by which cells globally inhibit translation initiation [10]. While significant attention has been paid to translational control at the elongation stage under stress, initiation control, and elongation both rely on the action of the target of rapamycin (TOR). Under mTORC1 stress, direct phosphorylation of eIF2α and ribosomal S6 kinase, along with eIF4E dephosphorylation and sequestration by binding proteins (4EBPs), are known to regulate global mRNA translation. However, these regulatory pathways alone do not entirely account for the observed degree of translational repression. The concept of translation reprogramming aligns with the regulation of translation during stress, enabling specific mRNA translation to maintain stress protein expression while halting overall protein synthesis. The regulatory processes can either affect global translation or the translation of specific sequences, and both will be treated independently in the manuscript. The manuscript could be split more obviously into **1**) global regulation; and **2**) selective regulation.

This review provides an overview of how the cellular translational machinery of higher eukaryotic (Human/mammalian systems, Yeast and Plants) responds to stress conditions and delves into the forms of translational modification induced by stress. It emphasizes the mechanisms of major signaling pathways involved in translational regulation during various stress types. The focus spans from well-established translational reprogramming in stress responses to recent advancements in understanding initiation and elongation modes of regulatory mechanisms. Additionally, we underscore the impact of translational control on cellular proteostasis, emphasizing processes that could potentially disrupt the aging process.

## 1. Overview of the Eukaryotic Translation Process

The translation of messenger RNA (mRNA) constitutes a complex, energy-demanding process characterized by multiple steps and factors [11,12]. It involves the decoding of triplet nucleotide codes on mRNA by tRNA and ribosomes to synthesize polypeptide chains. Translation initiation necessitates coordinated interactions between eukaryotic mRNAs and translation initiation factors, along with the 40S ribosomal subunit [13,14]. The m7GpppN cap structure, present on nuclear-transcribed cytoplasmic eukaryotic mRNAs, plays a critical role in splicing, mRNA stability, polyadenylation, and translation. This cap structure's association with eIF4F is an initial step in mRNA translation, comprising three subunits: eIF4A (a DEAD-box RNA helicase), eIF4E (a cap-binding protein), and eIF4G (a molecular platform with multiple docking sites). The primary function of the eIF4F complex is to recruit ribosomes to mRNA molecules.

mRNA translation involves four stages: initiation, elongation, termination, and ribosome recycling [15]. Initiation aims to prime the 40S ribosomal subunit for mRNA binding, recruit it to the mRNA, and position it at the initiation codon before the joining of the 40S complex with the 60S ribosomal subunit. Recruitment of the 40S ribosome subunit during translation initiation is deemed crucial [14]. This intricate process, involving numerous initiation factors and accessory proteins, is considered the rate-limiting step in mRNA translation [16]. Protein-RNA and protein-protein interactions play pivotal roles in translation initiation [17,18]. Notably, among the twelve distinct factors involved in eukaryotic translation initiation, DEAD-box RNA helicases, eIF4A, and Ded1p (DDX3 in humans) are highly conserved [7,19,20]. The interplay of eIF4A with eIF4E (m7G-cap-binding protein) and eF4G (a scaffolding protein) at the 5′-end of mRNAs is vital [14,19,20]. eIF4G orchestrates initiation by facilitating the recruitment of additional factors, providing a scaffold for ribosome/mRNA bridging [21]. eIF4A helicase unwinds RNA duplexes in vitro [22] and stimulates ATP-dependent RNA unwinding activity in interaction with eIF4G's MIF4G domain [23,24]. The basic translation initiation process is shown in **Fig 1.**

Transcription and translation guarantee the precise transfer and placement of genetic information into folded, functional proteins, ensuring their correct positioning within a cell for optimal functionality. Cellular protein synthesis is intricately regulated to match intracellular demands and external conditions. This process aids in minimizing the overall cost of protein production. Among cellular functions, mRNA translation stands out as a notably energy-intensive process, demanding approximately 75% of the cell's total energy [25]. Reduced mRNA translation increases cellular resource availability, redirecting resources toward cellular maintenance and repair [26]. Quality control mechanisms to prevent the synthesis of defective proteins, including cotranslational mRNA and protein surveillance at the ribosome [16].

Under stressful conditions, cellular pathways regulating ribosome biogenesis and mRNA translation play critical roles in stress survival. Aberrant proteins synthesized during stressful conditions undergo degradation pathways [27,28]. Failure in stress pathways can lead to the accumulation of misfolded proteins, potentially causing various human diseases [29-33] such as Alzheimer's, Huntington's, and Parkinson's diseases [34, 35]. Depletion or inhibition of translation components can reduce the risk of toxic protein accumulation, potentially extending lifespan [36]. Inhibition of mTOR signaling, a key regulator of translation initiation, has shown increased lifespans in various organisms, indicating its essential role as a longevity regulator [37-39].

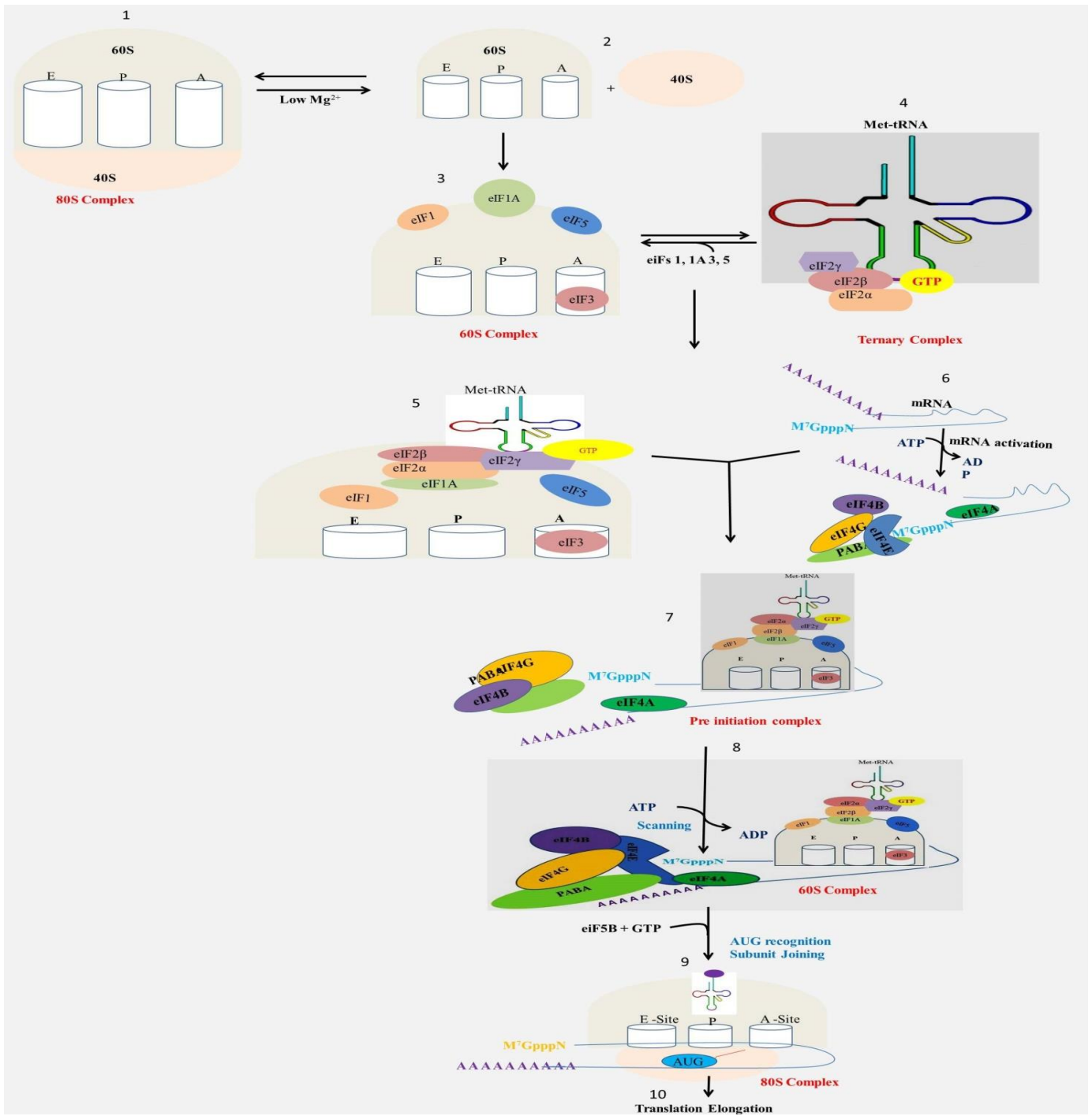

**Figure 1: Eukaryotic Cap-Dependent Translation Initiation and its Key Regulatory Pathways**: Eukaryotic mRNAs comprise a 5' m7G cap, which is bound by the eukaryotic initiation factor 4F complex (eIF4E, eIF4G, and eIF4A), and the ternary complex (eIF2-GTP-Met-tRNAi). Translation initiation starts with the assembly of the 43S preinitiation complex (PIC), consisting of the 40S ribosomal subunit, the ternary complex, and the initiation factors (eIF1, eIF1A, eIF3, eIF5). The PIC is recruited to the 5' cap of the mRNA by the eIF4F complex and eIF4B. Binding of eIF4F to the 5' cap and PABP to its poly(A) tail activates the mRNA. Successively, the 48S initiation

complex is formed, and TC delivers Met-tRNA into the P-site of the ribosome. Before joining of the 60S ribosomal subunit to the PIC, all initiation factors are released from the 40S small ribosomal subunit. Finally, eIF5B unites the 40S and 60S ribosomal subunits to form the 80S initiation complex, and translation elongation begins.

## 2. Translational reprogramming under stress

The regulation of gene expression and translation is critical and serves as a fundamental mechanism influencing various cellular processes, including growth, differentiation, and survival. Precise control of transcription and translation is required to maintain optimal levels of essential proteins for correct protein homeostasis. During adverse stress conditions, cells need rapid and efficient changes in mRNA translation. Stress-responsive protein synthesis targets and reprograms global translation and translation of specific mRNA [16]. While variations in mRNA expression levels play an obvious role in determining the cellular level of proteins, several studies indicate a lack of a consistent relationship between these phenomena. Unique abundant mRNAs may exhibit poor translation, and vice versa [40-42]. Alterations in translational machinery components or their availability can lead to mRNA translation arrest. Additionally, subcellular changes in the localization of messenger ribonucleoproteins (mRNPs) significantly impact translation regulation [reviewed by 43].

Different phases of translation possess specific regulatory targets. The primary targets of translational control are the initiation factors eIF2 and eIF4E. Disruption of these factors influences most of the translatome and results in rapid and robust implications, which are characteristic of the integrated stress response [reviewed by 43].Phosphorylation of eIF2 globally regulates translation by inhibiting eIF2B recycling and its ability to bind Met-tRNAi, an interaction involved in the translation of nearly all mRNAs [44,45]. Signal transduction pathways, responsive to stresses and physiological stimuli, regulate the functions of eIF2 or eIF4E translation initiation factors. Stress-induced changes in translation initiation are reversible mechanisms that restore translation rates after stress withdrawal [reviewed by 43].

In mammals, eIF2α activation occurs via four different protein kinases, leading to an increase in phosphorylated eIF2α (p-eIF2α) during stress, subsequently inhibiting protein synthesis [46]. In order to redesign its proteome to change gene expression and renovate the essential signaling pathways to control stress and promote cell survival, cell reduce the rate of global translation [46]. In order to redesign its proteome to change gene expression and renovate the essential signaling pathways to control stress and promote cell survival, cells reduce the rate of global translation. Elevated p-eIF2α levels are expected to increase the translation of specific mRNAs encoding pro-survival and stress-responsive proteins, while potentially inhibiting mRNAs encoding housekeeping genes. ATF4, a key integrative stress response (ISR) gene, preferentially translates under stress. ATF4 mRNA contains upstream ORFs (uORFs) in its 5'-UTR, significantly influencing translation efficiency in response to p-eIF2α levels [47-49]. Inhibition of translation initiation and polysome disassembly cause the accumulation of untranslated mRNPs in the cytosol [50]. These untranslated mRNPs aggregate with various proteins [51], subsequently; arrested mRNPs condense into non-membrane-bound subcellular compartments termed cytoplasmic stress granules (SGs) [52]. Cellular stress, like heat shock and oxidative stress, triggers the formation of SGs through eIF2 phosphorylation [50] SGs exist in a dynamic equilibrium with polysomes [53]. Notably, the decrease in SG disassembly perpetuates the translationally arrested state of mRNPs [54].

### 3. tRNAs fine-tune translation under stress

Numerous positions bear conserved modifications within the long 70–90 nucleotides of tRNA, contributing to the structural landscape.. Among these, specific alterations within the anticodon loop of tRNAs assume pivotal roles in the reprogramming of translation, especially under stress conditions [55]. The kinetics of codon-anticodon interactions, altering transcript expression and stability, are significantly affected by Wobble base modifications [55,56]. Predominantly, methylation catalyzed by tRNA methyltransferases at the wobble nucleotide bases or ribose sugar represents the most prevalent tRNA modification [57]. Notably, methylation at the wobble base within the tRNA anticodon loop plays a significant role in preserving translational fidelity [58,59].

In response to cellular stress, cells exert control over the methylation and demethylation processes of tRNA bases, consequently fine-tuning translational accuracy [60]. Despite being initially considered degradation remnants, recent research has shed light on the role of cleaved tRNA fragments in modulating protein synthesis during cellular stress, proliferation, and differentiation [61,62].The Ribonuclease angiogenin catalyzes stress-dependent cleavage of tRNAs, specifically targeting the anticodon loop, producing tRNA halves termed tRNA-derived stress-induced RNAs (tiRNAs).

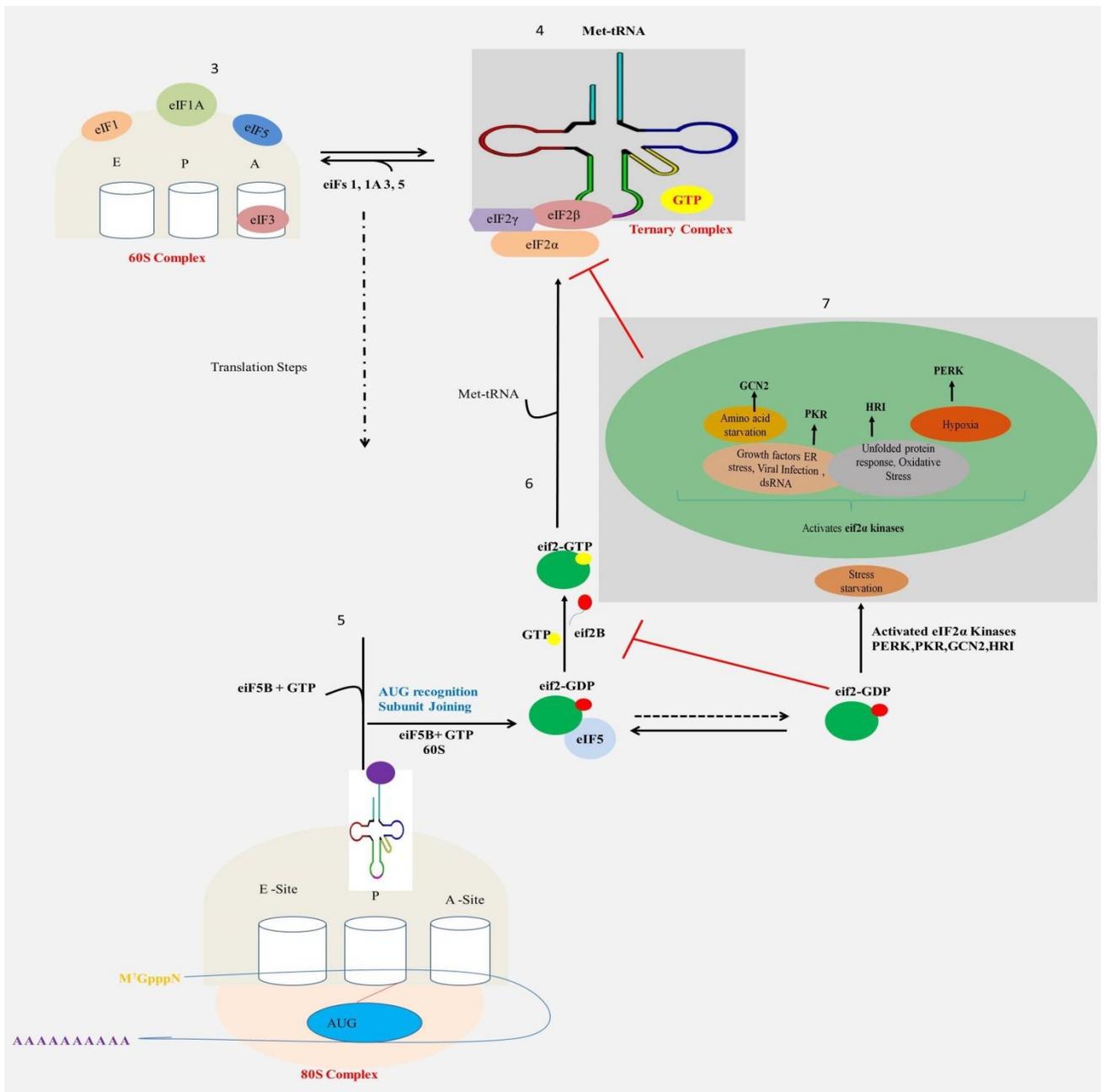

**Figure 2: Translation Regulation by eIF2α** eukaryotic initiation factor 2 alpha is a pivotal factor in the initiation of translation. When eIF2α undergoes phosphorylation, it impedes the assembly of the ternary complex (composed of eIF2, GTP, and initiator tRNA), thereby hindering translation initiation. The eIF2a kinases serve as rapid responders to disruptions in cellular equilibrium. This family comprises four members: PKR-like ER kinase (PERK), double-stranded RNA-dependent protein kinase (PKR), heme-regulated eIF2a kinase (HRI), and general control nonderepressible 2 (GCN2). Each kinase is triggered by specific environmental or physiological stresses, reflecting their distinct regulatory pathways. PERK, PKR, HRI, and GCN2 kinases are activated by signals such as ER stress, viral infection, and other cellular stressors,

leading to the phosphorylation of eIF2 α, a central component of the integrated stress response. Consequently, there is a global attenuation of cap-dependent translation.

Underlying this intricate network of regulation is the amino acid control (GAAC) pathway, which monitors and responds to stress-induced nutrient scarcity. Accumulation of noncharged tRNAs during nutrient limitations triggers the activation of the GCN-2 gene [63]. Activated GCN-2, stimulated by binding noncharged tRNAs, orchestrates the response to restriction of nutrients [64]. Furthermore, activated GCN-2 facilitates eIF2α phosphorylation, leading to the inhibition of mRNA translation. Its activation also initiates the expression of genes involved in amino acid biogenesis pathways [63,65]. Studies in yeast and *C. elegans* have elucidated an interplay between the GCN-2 and TOR pathways. In yeast, inhibition of the TOR pathway activates GCN-2 [65,66], while in *C. elegans*, GCN-2 regulates the expression of PHA-4, a downstream transcription factor of LET-363 (the orthologue of TOR) [65]. Notably, GCN-2's role resembles that of the TOR pathway, contributing significantly to lifespan regulation; loss of GCN-2 function reduces lifespan during amino acid limitation [65]. Collectively, these findings underscore a close interconnection between the TOR and GCN-2 pathways, emphasizing how nutrient restrictions can trigger a sophisticated system governing mRNA translation and longevity regulation. This basic mechanism of regulation is shown in **Fig 2.**

### *4. mRNA translation regulation of by chaperones*

Hsp27, a key factor in heat shock stress response, plays a crucial role in mRNA translation regulation through interactions with eIF4G and PABP1 [67-70]. Under heat shock conditions, eIF4G separates from the PABP1-eIF4G complex and engages with Hsp27, prompting its translocation from the cytosol to the nucleus. This suggests that Hsp27 may facilitate the transfer of Eif4G into the nucleus [68]. Previous research has indicated that following dissociation from PABP1, the eIF4G-Hsp27 complex localizes in insoluble heat shock granules [67]. Recent studies have identified the Hsp27-eIF4G-PABP complex in the cytosol, suggesting that the binding of Hsp27 alone may not solely be responsible for mRNA translation attenuation [68]. At higher temperatures, the interactions between eIF4G and/or PABP1 with mRNA might be impeded, indicating a decoupling of mRNA nuclear export and translation. This phenomenon might elucidate the observed nuclear accumulation of eIF4G/PABP1 during stress conditions [68].This data underscores the impact of Heat Shock Proteins (HSPs) on mRNA translation during stress, highlighting their role in reducing new protein synthesis to alleviate the additional burden on chaperones.

## 5. Selective translational regulation of specific mRNAs during stress

The process of eukaryotic protein synthesis involves multiple stages, necessitating correspondingly an intricate regulatory system. This multi-stage mechanism is essential to tightly control the translation of critical proteins, as their misexpression could prove lethal to the cell. The association of mRNAs with the translation machinery can either be inhibited or stimulated, constituting a pivotal aspect of translational control. This regulatory landscape encompasses two primary categories: mRNA-specific regulation and global regulation [71]. mRNA-specific regulation selectively influences the translation of particular mRNAs, while global regulation modulates the overall translational efficiency of numerous mRNAs through generalized alterations in the translation process. Despite the documentation of diverse translational regulation processes, only a few are comprehensively understood mechanistically. Translation initiation typically serves as a rate-limiting step [15], wherein among the translation initiation factors, eIF4E levels are notably low. The formation of the eIF4F complex and subsequent translation initiation represent critical, rate-limiting phases in most circumstances. Extensive research efforts are dedicated to uncovering the molecular underpinnings of this singular regulatory stage. The binding of eIF4E to the 5' mRNA cap stands as a pivotal step. Additionally, the recruitment of mRNA to the 40S ribosomal subunit is also deemed a rate-limiting initiation step [13,16]. Other significant steps involved in regulating translational initiation encompass the availability of the ternary complex (TC) (via eIF2α phosphorylation) [72], modulation of mRNA poly(A) tail length—an enhancer of translation and mRNA stability [73-75]. Furthermore, the control of translation initiation via cap-independent mechanisms, as well as subsequent steps like initiator codon recognition and scanning, may also be subject to modulation [76,77]. Phosphorylation emerges as a key modifier of various initiation factors, often utilized to regulate global rates of protein synthesis. Beyond phosphorylation, a multitude of other posttranslational modifications—such as methylation, glycosylation, and ubiquitination—play crucial roles in translational regulation and warrant extensive study [16].

### 5.1 eIF4F-Mediated 5'-Cap Recognition Regulation

Under normal circumstances, mTOR organizes the assembly of the eIF4F complex at the 5' end of mRNA, facilitating the recruitment of ribosomal subunits and subsequent translation of the transcript [78]. The prevalent mechanism in eukaryotes for regulating translation initiation rates involves the eIF4F method of mRNA 5'-cap recognition. The stability and translational efficiency of eukaryotic mRNAs are notably influenced by the 5' end cap [79-81]. The presence of a methyl group on the 5'-cap is crucial for recognition by CBP, eIF4E, and the decapping enzyme (DcpS) [82,83]. Decapping serves to deactivate translation initiation and initiates the 5'-to-3' decay of mRNA [84]. Under certain conditions, mRNAs cannot be translated via cap-dependent translation. For instance, cap-dependent translation is inhibited during cellular stress and viral infection [8,85,86]. Approximately 10% of human mRNAs feature 5' UTRs that enable cap-independent translation initiation during stress [15, 87-90. Although the translation initiation factor eIF4F has been traditionally associated with cap-dependent translation, current research investigates stress-specific variations in eIF4F [91-93]. This mechanism of regulation has been illustrated in **Fig 3**.

### *5.2 eIF4E regulated ternary Complex formation*

A ternary complex is a ribonucleoprotein complex containing aminoacylated initiator methionine tRNA, GTP, and initiation factor 2 (eIF2 in eukaryotes, or IF2 in prokaryotes). In prokaryotes, the initiator is fMet-tRNA, while in eukaryotes, it is Met-tRNAi. In eukaryotes, the 43S ribosomal pre-initiation complex (PIC), consisting of the 40S ribosomal subunit and the eIF2-GTP-initiating Met-tRNAi ternary complex, along with additional eIFs such as eIF1, eIF1A, eIF3, and eIF5, is initially recruited to the 5' terminus of mRNAs. Subsequently, it scans the 5' untranslated region (5'UTR) before moving toward the start codon. Once the start codon is reached, the 60S ribosomal subunit associates with the complex, forming the 80S initiation complex. This complex then facilitates the recruitment of the correct aminoacyl-tRNA into the A (aminoacyl) site, initiating the synthesis of the first peptide bond and transitioning the initiation phase toward elongation [94] illustrated in **Fig 1.**

The translation initiation factors eIF4E plays a crucial role in facilitating the association of the eIF4F complex with the 5' cap structure of mRNA, serving as a significant rate-limiting factor in the initiation of canonical protein synthesis. Various proteins in animals exert strict control over eIF4E translational activity through phosphorylation or direct binding to eIF4E [95]. Among the well-known regulators of eIF4E translation are 4EBP [96] which bind to distinct lateral and dorsal sites of eIF4E [97-99]. The comprehensive mechanism of 4EBP translation initiation regulation has been illustrated in **Fig 3**. The interaction between 4EBPs and eIF4E closely correlates with their phosphorylation status, regulated by mTOR [100]. When in a hypophosphorylated state, 4EBP forms a tight complex with eIF4E. The phosphorylation of eIF4E at the Ser-209 residue by the kinases Mnk1 and Mnk2, triggered by tumor promoters, growth factors, and mitogens, significantly influences its oncogenic activity [95,101,102]. The phosphorylation of eIF4E has been shown to selectively regulate the translation of specific mRNAs encoding proliferation, pro-survival (such as BIRC2 and Mcl-1 mRNAs), angiogenesis-related proteins (e.g., VEGFC), and extracellular matrix proteins (MMP3, MMP9) [103]. Studies indicate that eIF4E phosphorylation enhances cellular tolerance to stress, including oxidative and cytotoxic stress, thereby promoting cell survival, regeneration, proliferation, and tumor progression. This process seems to operate through interactions with 4E-transporter (4E-T) protein and the qualitative control of protein synthesis [104]. It is imperative to conduct detailed research into the interactions between eIF4E phosphorylation, cellular stress, and survival. Furthermore, eIF4E also facilitates the nuclear-cytoplasmic export and degradation of specific mRNAs containing 50-nucleotide elements in their 3' UTR [65,105-107].

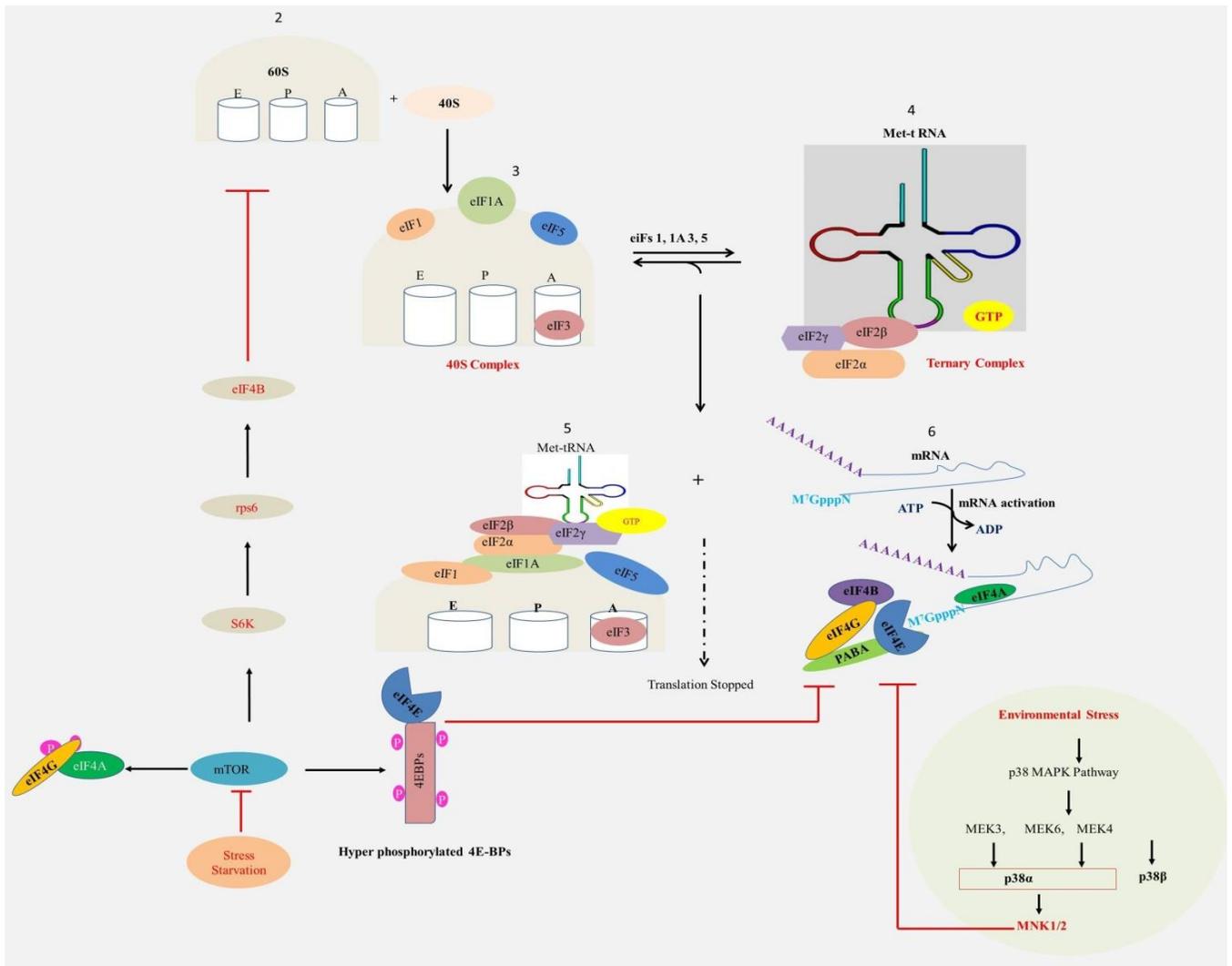

**Figure 3: Translation Regulation by mTOR , 4EBPS and** MNK pathway:The eIF4F complex plays a vital role in translation initiation, particularly in cap-dependent translation. It comprises three primary subunits: eIF4E: This protein binds to the 5' cap structure of mRNA, serving as the cap-binding protein. eIF4A: An ATP-dependent RNA helicase, eIF4A unwinds the secondary structure of mRNA. eIF4G: Acting as a scaffold protein, eIF4G facilitates interactions between eIF4E, eIF4A, and mRNA. Additionally, there's 4EBP (eIF4E-binding protein), which binds to eIF4E, preventing its interaction with eIF4G and thereby obstructing eIF4F complex formation. Under stress or starvation conditions, 4EBP binding to eIF4E inhibits translation initiation, while eIF2$\alpha$ phosphorylation reduces ternary complex formation. mTORC1 promotes the hyper-phosphorylation of 4EBP, preventing its inhibitory association with eIF4E, thus facilitating eIF4F complex assembly. Furthermore, mTOR enhances the phosphorylation of eIF4G and eIF4B, either directly or via S6 kinases. Given that eIF4E is the most limiting subunit of the eIF4F complex, its availability is crucial for recruiting eIF4A to mRNA. eIF4E activity is regulated by MAPK pathways, which directly phosphorylate eIF4E via the MNK protein kinases.

Proteins such as LRPPRC, PRH, or 4E-T interact with eIF4E through the canonical eIF4E-binding motif [101,108,109], underscoring the critical role of this domain in eIF4E binding and regulation. Both eIF4E and 4E-T are components of stress granules (SGs) and processing bodies (PBs), yet the role of phosphorylated eIF4E in these intracellular structures remains insufficiently studied. Furthermore, other proteins serving as 4E-interacting partners establish connections between eIF4E and 4E-binding proteins via canonical eIF4E-binding motifs or related structures [110]. For instance, *Xenopus* Maskin and *Drosophila* Cup act as mRNA-specific 4EBPs [111], and in the nervous system, Neuroguidin is identified as a 4EBP [112]. These specific eIF4E interacting partners play crucial roles in regulating animal development programs, primarily by facilitating various protein-protein interactions that link the 5' and 3' UTRs of specific mRNAs, rendering them translationally inactive [113,114].

### 5.3. eIF2 and regulation of translation initiation

Accurate translation initiation is a crucial process for cells to synthesize the correct proteins [7,78]. The eIF2 protein plays a pivotal role in protein synthesis by binding to initiator tRNA (tRNAi Met) within the cytoplasm and transporting it to ribosomes, where it recognizes the AUG codon on mRNA. The TC complex (eIF2–GTP–tRNAi Met) serves as a critical intermediary in the translation initiation pathway. Comprising three subunits ($\alpha$, $\beta$, and $\gamma$), eIF2 is a heterotrimeric protein. The eIF2$\gamma$ core subunit interacts with GDP/GTP and tRNAi Met, while eIF2$\alpha$ facilitates AUG codon recognition and tRNAi Met-binding to the subunit. EIF2$\beta$ interacts with eIF2 ligands and other factors essential for the regulation of eIF2 function [115-118].

Extensive research has explored the inhibition of protein synthesis through eIF2$\alpha$ phosphorylation. During stress, phosphorylation of the eIF2$\alpha$ subunit impedes eIF2$\beta$'s ability to exchange GDP for GTP. Consequently, the formation of active TC is significantly reduced, leading to global translation downregulation [4,71]. Phosphorylated eIF2α–GDP competitively inhibits eIF2β, as it exhibits a higher affinity for eIF2β compared to unphosphorylated eIF2α–GDP, elucidating the molecular mechanism behind this inhibition [119]. Four types of kinases phosphorylate the conserved Ser51 residue of the eIF2α subunit in response to diverse stresses [115,120]. In eukaryotes, phosphorylation of the eIF2α subunit during stress constitutes a primary mechanism for regulating mRNA translation. In yeast, GCN2 kinase phosphorylates the eIF2α subunit under nutrient starvation conditions [63,121]. Similarly, during nutrient deprivation, protein misfolding, or immune responses in mammals, GCN2 phosphorylates eIF2α [122-124]. The global translational response to stresses involves a critical mechanism of eIF2-GTP regeneration [125]. Recently, a fail-safe regulatory switch has been identified, wherein eIF2β binds to and inhibits phosphorylated TC and TC/eIF5 complexes, offering an alternative pathway to inactivate eIF2/eIF2β complexes [10].This mechanism of translation initiation regulation has been illustrated in **Fig 2**.

## *6. Translation Elongation regulation and signalling pathways*

As discussed earlier, the initiation phase has been considered the predominant target for translational regulatory mechanisms. However, accumulating evidence suggests that control can also be exerted at the elongation and termination phases [126]. One common regulatory mechanism involves the phosphorylation of elongation factors, similar to some initiation factors.

Phosphorylation of the elongation factor eEF2 on the Thr56 amino acid residue within the GTP-binding domain is a well-known regulatory mechanism in response to oxidative stress.

This phosphorylation alters its affinity for the ribosome, leading to its inactivation [127-130]. Various stress signals activate eEF2 kinase (eEF2K) by AMPK-mediated phosphorylation on the serine 398 residue. eEF2K is a member of the α-kinase family, and its activation or inactivation is regulated by anabolic and mitogenic signaling pathways, such as mTORC1 and the MAP kinase pathway [131,132]. To facilitate the rapid continuation of translation elongation, eEF2K is degraded by the ubiquitin-proteasome system. The regulation of translation elongation by mTOR signaling at eEF2 has been extensively reviewed [133,134]. Additionally, eEF2 activity can be modulated by polyadenylation element binding protein 2 (CPEB2), an RNA-binding protein that reduces the GTP hydrolysis of eEF2 [135]. The mechanism of translation elongation regulation has been illustrated in **Fig 4**.

Under normal conditions, CPEB2, through binding to the 3'UTR, reduces the translation of HIF1α mRNA. Interestingly, CPEB2 dissociates from HIF-1α mRNA under hypoxic conditions, allowing for the synthesis of HIF-1α and adaptation to hypoxia. Furthermore, eEF2 activation is suppressed during amino acid starvation stresses, endoplasmic reticulum stress, energy stress caused by hypoxia, and genotoxic stress [130,136-138]. Uncharged tRNA accumulation can stall elongating ribosomes in these conditions and play a role in activating eIF2 kinase Gcn2 [139]. Recently, the binding of modified or damaged tRNA has been revealed during stress, contributing to both of these processes [140].

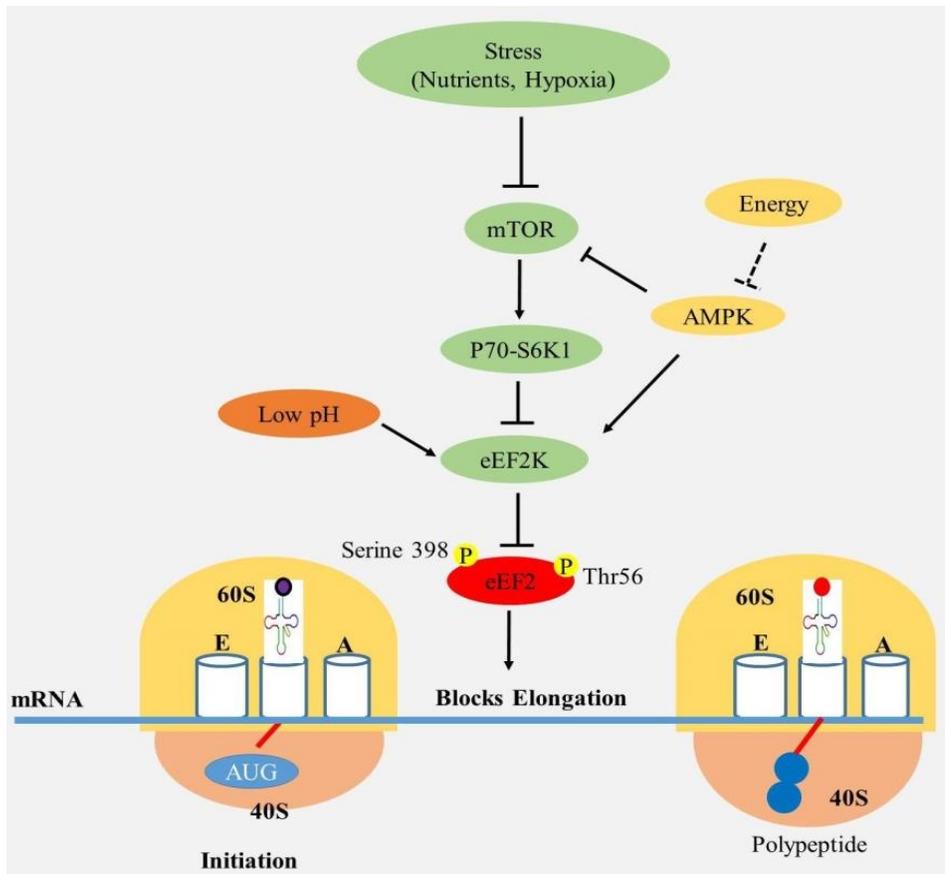

**Figure 4: Schematic overview of the regulation of mRNA translation elongation.** Upon 80S complex formation, the ribosome is primed for translation elongation. eEF2 is regulated by the mTORC1 and MAP kinase signaling pathways by controlling eEF2K during stress. eEF2K is activated or inactivated by phosphorylation at Thr56 and Serine 398 residues by anabolic and mitogenic signaling agents (mTORC1 and MAP kinase pathway). Acidic conditions inhibit the phosphorylation reaction catalyzed by eEF2 kinase (eEF2K) and block translation.

## 7. Developing Concepts in Translational Regulation

Considerable attention has been directed towards the regulation of mRNA post-transcriptional modifications for over a decade. The stability and translation of target mRNAs are also influenced by the more recently identified processing bodies and small RNAs cytoplasmic processing bodies and small RNAs. Herein, we expound upon these novel concepts, emphasizing their role in translational regulation. Furthermore, we examine recent examples illustrating how the modulation of alternative transcripts can impact translation.

### 7.1. P-bodies and translation regulation mechanisms

Processing bodies (P-bodies) and stress granules (SGs) are cytoplasmic ribonucleoprotein (RNP) granules that primarily constitute pools of translationally repressed mRNAs and proteins associated with mRNA decay, thereby indicating their roles in transcriptional regulation [141,142]. P-body mRNPs consist of translation repression factors i.e GW182, Pat1, Dhh1/RCK/p54, Caf1/Pop2, Edc3, Xrn1 and eIF4E and eIF4G [143], while SG mRNPs encompass a subgroup of translation initiation factors, potentially poised for reentry into translation. Notably, stress and genetic variations dictate the composition of P-bodies, SGs, and related RNA granules. These cytoplasmic granules are highly dynamic, with their formation and disassembly [144,145]. They are largely conserved in all eukaryotic cells and exhibit characteristics of liquid droplets. However, the precise role of these granules in translational repression or mRNA decay remains inadequately resolved. The specific mechanism governing the transport of these mRNAs into P-bodies for translational repression remains unidentified. Phase transition techniques have been employed to investigate how soluble RNPs condense into liquid or solid bodies [146-148]. These RNP bodies functionally regulate RNP exchanges, subcellular localization, and concentration [149,150]. Various pieces of evidence indicate a connection between P-bodies and translation. Under environmental conditions, mRNAs exist in two distinct states: associated with polysomes, actively translated, or associated with P-bodies, translationally repressed P-bodies' size and number vary depending on cellular conditions, such as glucose depletion, osmotic stress, UV light, and acid stress [151,152].

Amino acid deprivation has been demonstrated to induce the formation of P-bodies in mammalian cells, along with the localization of mRNA within these structures. This evidence strongly supports the idea that the translation rate has a direct impact on the number of P-bodies. Initially, P-bodies were considered sites for mRNA decay, primarily due to their association with decapping complexes [153]. P-bodies are increasingly recognized as temporary repositories for translationally silenced mRNAs that, without undergoing decay, can reenter the translating pool [154-156] particularly in processes like early development, oogenesis, and neuron plasticity [157]. Several studies suggest that translation repressors are essential, but decay machinery is not, for P-body formation [158]. P-bodies can be disrupted without halting decay [159,160]. Although P-bodies might serve as centers for controlling mRNA destiny [161], the identity of P-body mRNAs and their specific role remain poorly defined.

### 7.2. Stress granules and translational regulation mechanisms

Cytoplasmic stress granules (SGs) intricately intertwine with mRNA translational regulatory pathways during periods of stress [162]. SGs are delineated as stress-induced membraneless, phase-dense cytoplasmic bio-condensates comprising translationally silent mRNAs associated with preinitiation complexes i.e RNA translation initiation factors (eIFs), and RNA-binding proteins (RBPs), precipitating assembly when translation initiation is obstructed [163]. Notably, SGs lack eIF5 and eIF2, essential proteins for the transformation of ribosomal preinitiation complexes into translationally competent ribosomes [164]. Following the resolution of stress, SGs can transition back into polysomes [165-167]. SGs are believed to form through liquid-liquid phase separation, driven by cooperative interactions among RNA-RNA, protein-RNA, and protein-protein interactions [168-170]. Despite analogies to P-bodies and some common components, SGs contain specific constituents, including translation initiation factors, 40S ribosomal subunits, and Antioxidant Response Element (ARE) binding proteins [54]. However, fusion events and close associations between SGs and P-bodies are evident [54,171]. Formation of stress granules has been illustrated in **Fig 5**.

Stress-induced eIF2α phosphorylation is both essential and sufficient for SG assembly [53, 167]. As discussed earlier, stress-induced phosphorylated eIF2α inhibits eIF2B activity, halting translation initiation and promoting SG assembly [9,45]. Under stress, various transcripts that escape eIF2α phosphorylation-mediated translational arrest produce proteins protecting cells from stress-induced damage. For instance, HSP70 mRNAs are selectively translated and excluded from SGs [167,172]. The Integrated Stress Response (ISR) kinase family, associated with eIF2α phosphorylation during stress, senses charged tRNAs through GCN2 [173], dsRNA through PKR [174], ER stress through protein kinase-like ER kinase (PERK) [124], and redox state and heme stress through HRI [175]. Deletion or inactivation of these eIF2α kinases renders cells insensitive to the respective stresses, with each kinase triggered by distinct stress types, subsequently leading to SG formation [reviewed in 123]. Following SG assembly, pharmacological manipulations, such as the PERK signaling inhibitor ISRIB (Integrated Stress Response Inhibitor), can reverse eIF2α phosphorylation by activating eIF2B. ISRIB treatment has been shown to reverse SG assembly and readily restore translation [176, 177].

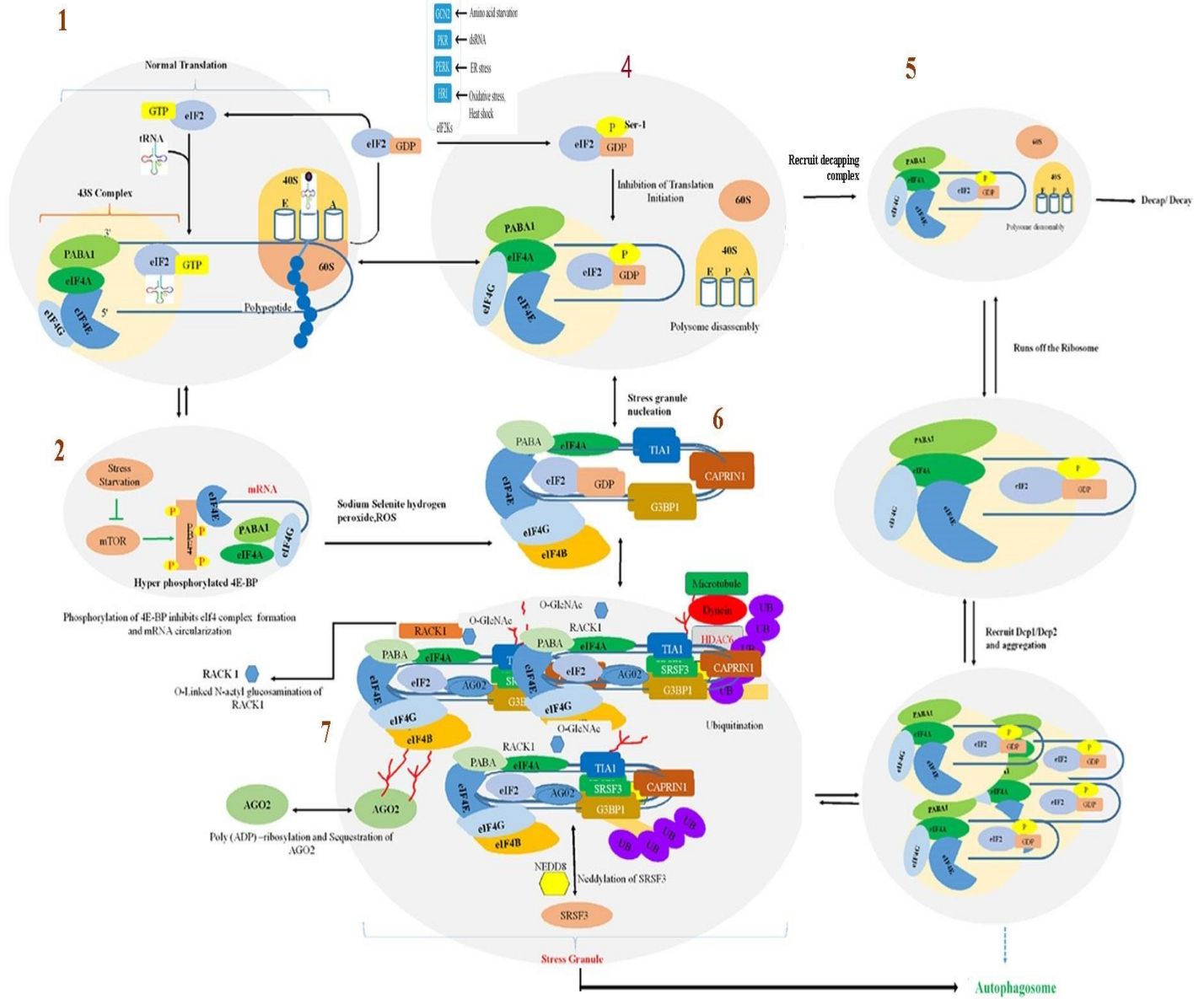

**Figure 5: Overview of translation regulation and stress granule formation:**

Transcribed RNAs form nuclear messenger ribonucleoprotein particles (mRNPs). RNA-binding proteins linked with mRNA are shifted to the cytosol, where they govern cytoplasmic localization and translational proficiency of the mRNA. mRNA conversion into proteins activates the assembly of translation complexes, and when this process is stalled, the mRNPs accumulate as stress granules.

Step 1: In normal translation, the eIF4F complex recruits the 43S ribosomal subunit. Upon recognition of the initiation codon by the anticodon of tRNAMet, eIF2-GTP is hydrolyzed, and eIF2-GDP is released, and early initiation factors are displaced by the 60S ribosomal subunit.

Step 2: Under stress, mTORC1 promotes the hyper-phosphorylation of 4EBP, inhibiting the association of eIF4E with eIF4G and eIF4A to form the eIF4F complex properly, and blocking translation.

Step 3: In stressed cells, phosphorylation of eIF2α by GCN2, HRI, dsRNA, PERK, and/or PKR converts eIF2 into a competitive antagonist of eIF2B, depleting the stores of eIF2/GTP/tRNAMet. This stops the exchange of GDP-GTP and the restoration of the 43S pre-initiation complex, inhibiting translation.

Step 4: Upon eIF2α subunit blockage, elongating ribosomes 'run-off' the mRNA.

Step 5: Following translation, mRNAs can quit translation and assemble a translationally repressed mRNP that can be either degraded or assembled into P bodies. mRNAs enclosed by P bodies can be subject to decapping and 5'-3' degradation, or they can exchange P-body components for stress granule components to revert translation. Before re-entering translation, mRNAs should obtain extra translational components (eIF2, eIF3, and 40S subunits). Specific factors like mRNA binding proteins or the presence of a poly(A) tail might affect this process. So, mRNPs within stress granules can return to translation initiation again and enter polysomes or can be targeted for autophagy.

Step 6: Translation inhibition, which causes the formation and aggregation of stress granules, is a consequence of the marked accumulation of untranslated mRNPs as a result of blocked initiation of translation, and the formation of stress granules themselves is not required for translation arrest. GTPase-activating protein-binding protein 1 (G3BP1) and T cell-restricted intracellular antigen 1 (TIA1) attach to the polysome-free mRNAs and build up to nucleate stress granule formation. Sodium selenite or hydrogen peroxide treatment (ROS generation) inhibits the functions of mTOR, resulting in stress granule formation. For granulation, G3BP1 must be dephosphorylated and demethylated. Poly(ADP)-ribosylated stimulates stress granule nucleation. Binding G3BP1 to cell cycle-associated protein 1 (CAPRIN1) encourages the formation of stress granules.

Step 7: Aggregate large stress granules from smaller focal points. This approach includes retrograde microtubule-dependent trafficking mediated by dynein motors and histone deacetylase 6 (HDAC6) binding to G3BP1, microtubules, and polyubiquitin chains enriched in stress granules. Post-translational modifications of stress such as poly(ADP)-ribosylation and O-linked N-acetylglucosamination (O-GlcNAc) control the recruitment of different proteins. The O-GlcNAc-dependent recruitment of a receptor for activated RACK1 protein to stress granules causes RACK1-mediated pro-apoptotic signaling to be sequestered and inhibited. Neddylation, a post-translational modification, is necessary for the formation of stress granules by mediating the serine/arginine-rich splicing factor 3 (SRSF3) interactions with the eIF4F complex, N-acetylglucosamine, mRNP, messenger ribonucleoprotein.

### *7.3. miRNAs and Translational Regulation mechanism*

In various biological pathways, two types of small RNAs, microRNAs (miRNAs), and short interfering RNAs (siRNAs) have emerged as significant gene regulators. These non-coding regulatory RNAs promote the stability and translation of mRNA, resulting in repressed gene expression [180] [178] . Both miRNAs and siRNAs are approximately 21-26 nucleotides in length, and their differentiation is based on their biogenesis [181-183] [179-181]. These miRNAs are derived from precursors of more than 70 nucleotides, which are hairpin segments with poor base-pairing. In contrast, siRNAs originate from RNA precursors that are perfectly complementary, either post-transcriptionally or co-transcriptionally [182]. The biogenesis of siRNAs is categorized into canonical and non-canonical pathways [183]. Most miRNAs in animals are processed by the consecutive action of the RNase III-like enzymes Drosha and Dicer from longer hairpin transcripts, whereas in plants, only Dicer is involved in this process. The Argonaute family protein (AGO) is loaded onto one strand of the hairpin duplex [183] (forming the core of silencing complexes (miRISCs) induced by miRNA. This complex can silence the expression of target genes at the post-transcriptional level. Bioinformatics studies suggest that the human genome encodes approximately 1000 miRNAs, each regulating about 10 mRNAs, potentially affecting over 30% of all genes [184, reviewed by 185].

miRNAs are known to interact with both the 5' and 3' untranslated regions of target mRNAs, inducing mRNA degradation and translational repression [182, 186,187]. Recent studies indicate that miRNAs localize to multiple subcellular compartments to regulate translation and transcription rates [188]. Remarkably, all subcellular sections involved in miRNA-mediated mRNA expression repression seem to concentrate in P-bodies. P-bodies are believed to be sites where mRNAs are stored and occasionally degraded away from the translation machinery [189]. P-bodies contain 5' and 3' exonucleases, decapping enzymes, and Ago proteins (GW182 and Rck/p54), suggesting that P-bodies play a role in miRNA-mediated repression or degrade mRNA machinery [141,190]. Cytoplasmic stress granules also contain

Ago proteins, miRNAs, and their target mRNAs [191]. The localization of Ago proteins in stress granules and P-bodies is miRNA-dependent, suggesting a potential collaboration between stress granules and P-bodies in miRNA-mediated translation regulation [190].

The mechanism used by miRNAs to regulate target gene expression has been a controversial subject. *In vivo* and *in vitro* studies confirm that miRNAs, depending on various factors, can inhibit translation, destabilize mRNA, or both [192]. In both animals and plants, miRNAs can silence targets through RNA degradation as well as translational repression pathways [192]. In animals, binding of miRNA-induced silencing complexes (miRISCs), containing GW182 proteins, to 3'UTR target sequences recruits deadenylation factors, eliminating the poly(A) tail and making the mRNA susceptible to exonucleolytic degradation [193-196].In plants, a common mechanism involves the perfect pairing of miRNA with its target site, supporting endonucleolytic cleavage of mRNA by Argonaute [197-199]. Both in animals and plants, there are instances where miRNAs cause reduced protein (but not mRNA) levels, suggesting that translational repression is directed by miRISC. Recent studies indicate that the Carbon Catabolite Repression—Negative On TATA-less (CCR4–NOT complex), recruited by GW182, deadenylates target mRNAs, resulting in the repression of translation initiation [200-204]. The exact mechanism causing the inhibition of protein production is not clear, but in animals, it has been proposed to occur at initiation, elongation, co-translational protein degradation, and premature termination of translation [159,185,203,204]. There is also increasing evidence suggesting that miRNAs interfere with the functions of the IF4F complex and PABPC during translation and/or mRNA stabilization [73]. From the discovery of miRNAs, significant progress has been made in understanding how cells produce miRNAs, their regulatory effects on the central dogma, and their involvement in various physiological and pathological events. Recent studies have highlighted that miRNAs can serve not only as biomarkers for diseases but also play a significant role in intercellular communication [205-209]. However, our knowledge regarding when and how miRNAs exert regulatory effects on transcription is still insufficient. Additionally, conditions under which miRNAs cause translational activation need further exploration.

## *8. Translation Regulation by mTOR through 4EBP and S6K1*

Protein translation regulation primarily occurs during two pivotal stages of translation initiation: the recruitment of the 40S ribosomal subunit and the loading of initiator methionyl-tRNA onto the 40S ribosomal subunit [71,210]. The translation initiation factor eIF4E oversees 40S subunit recruitment and is inhibited by the binding proteins 4EBP [111,211]. The 4EBP is an important regulator of overall translation levels in cells. Meanwhile, the eIF2α protein governs tRNA loading, controlling the recycling of eIF2 [71]. mTOR, a member of the phosphatidylinositol kinase family, forms two distinct complexes, mTORC1 and mTORC2 [212]. Reduced mTOR signaling under stress inhibits processes such as Ribosomal protein (r-protein synthesis), Ribosomal DNA (rDNA) transcription, and mRNA translation initiation [213-214]. mTORC1 primarily promotes protein synthesis by phosphorylating two key effectors, eIF4E Binding Protein (4EBP) and p70S6 Kinase 1 (S6K1). This phosphorylation prevents the assembly of the eIF4F complex by 4EBP, which directly interacts with eIF4E [44,216]. Under optimal conditions, mTORC1 constitutively phosphorylates 4EBP at multiple sites and their phosphorylated variants (p-4EBPs) cannot bind eIF4E. mTORC1 phosphorylation trigger its dissociation from eIF4E [122,194]. During stress, mTORC1 inactivation leads to dephosphorylation of p-4EBPs. These dephosphorylated 4EBP variants bind to eIF4E, inhibiting eIF4F complex assembly and subsequently leading to translation inhibition [215]. The binding sites of 4EBPs and eIF4G on eIF4E overlap.

Another class of proteins governing translation during stress is S6 kinases (S6Ks) [217]. Under optimal conditions, mTORC1 directly phosphorylates S6K1, enabling its activation by PDK1 [218]. Phosphorylated S6K1 targets ribosomal protein S6 (RPS6), a component of eIF4B and the 40S ribosomal subunit, which promotes eIF4A's helicase activity [217,219,220]. The role of phosphorylated S6K in facilitating translation remains unclear, although its inactivation by mTORC1 during stress is predicted to inhibit translation [reviewed by 221]. S6K1 phosphorylates and activates various substrates, including eIF4B, a positive regulator of the eIF4F 5' cap binding complex [218,222]. Moreover, S6K1 degrades the eIF4B inhibitor PDCD4 through phosphorylation [223,224]. Interestingly, the binding of mTORC1 with eIF3 promotes interactions between the 40S subunit and eIF4G, enhancing the recruitment of 40S ribosomal subunits to the eIF4F cap-bound complex and facilitating protein initiation complex (PIC) assembly [225]. This orchestrated network involving mTOR and its associated complexes, as well as various kinases and regulatory proteins, intricately governs translation initiation under different physiological conditions. The comprehensive mechanism has been illustrated in a **Fig 6**.

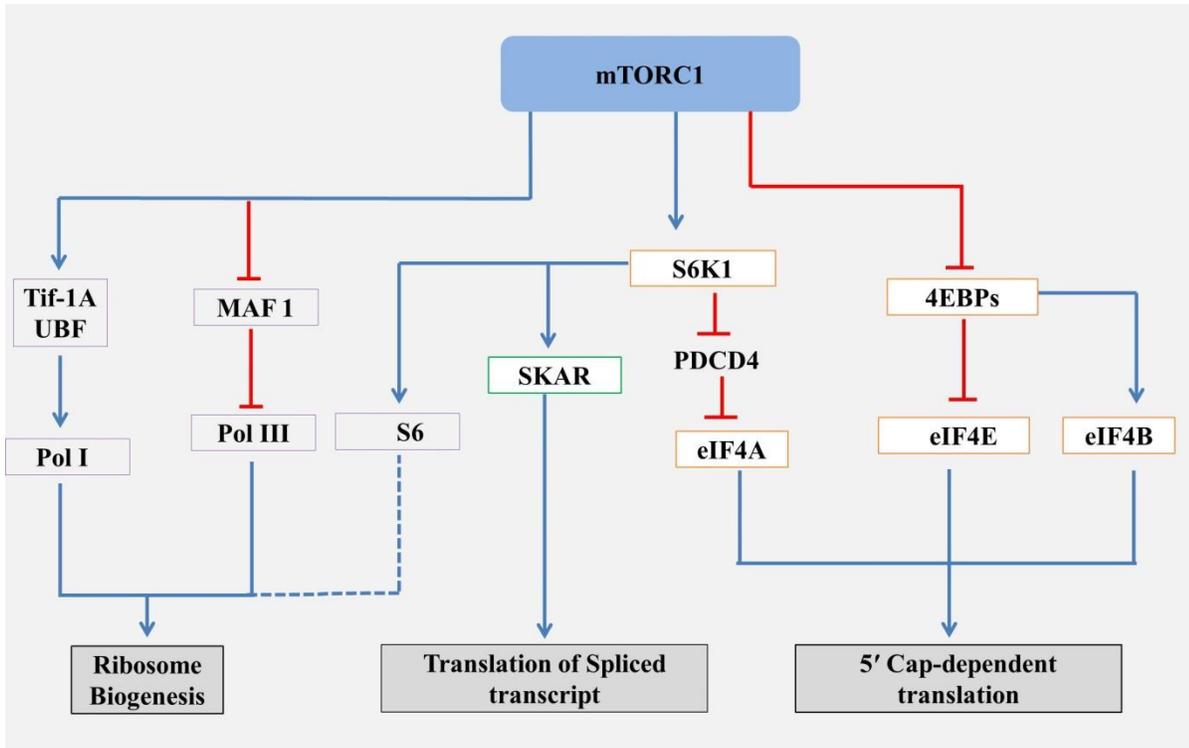

**Figure 6:** mTORC1 activation initiates downstream catabolic processes and inhibits autophagy and lysosome biogenesis, among other macromolecules, while enhancing anabolic programs such as the production of proteins, lipids, and nucleotides. mTORC1 inhibition affects mRNA translation, splicing, and ribosome biogenesis. mTORC1 regulates its activity through its substrates 4EBP2 and S6K1.

## *9. Ribosome biogenesis and translation regulation mechanisms*

mTORC1 modulates RNA polymerase III-dependent transcription by interacting with transcription factor-IIIC and Maf1[226]. Studies have revealed that mTOR directly interacts with ribosomal DNA promoters, leading to chromatin remodeling and subsequent activation of ribosomal gene transcription [227,228]. The growth-dependent transcription initiation factor (TIF-IA) promotes the activation of ribosomal DNA gene transcription by interacting with RNA polymerase I [229]. Inhibition of mTORC1 by rapamycin/stress leads to TIF-IA inactivation, resulting in the downregulation of 47S pre-rRNA transcription [214]. This comprehensive mechanism has been illustrated in **Fig 6**. Phosphorylation of eIF2α leads to the formation of stress granules (SGs), preferentially inhibiting the translation of mRNAs encoding ribosomal proteins [230]. The mTOR stress-response pathway primarily targets the translation inhibition of ribosomal protein mRNAs [212] and mTOR inactivation further impedes the translation of these mRNAs. Moreover, active mTOR aids in the transcription of RNA Pol I-dependent rRNA genes [231]. Thus, modifications in mTOR activity directly regulate both rRNA transcription and ribosomal protein mRNA translation. Ribosome biogenesis encompasses various steps occurring within three distinct subnucleolar components, starting from the initiation of Pol I transcription to the processing of pre-rRNA and eventual ribosomal assembly. Any disruption in this highly orchestrated process can lead to nucleolar stress [232], resulting in cellular insults.

Ribosome biogenesis is a highly energy-consuming process, with studies suggesting that approximately 75% of transcriptional activity is dedicated to this fundamental cellular function [233]. In response to stress, repression of ribosomal protein synthesis has been observed [234]. The orchestration of ribosome biogenesis is intricately regulated by both internal and external signaling mechanisms within the cytoplasm and mitochondria. The generation of ribosomes necessitates the involvement of all three RNA polymerases (I, II, and III). Within nucleoli, RNA polymerase I drives the transcription of the polycistronic 47S pre-rRNA, which upon splicing produces three distinct types of rRNAs (18S, 5.8S, and 28S). Meanwhile, RNA polymerase II and III act in the transcription of genes encoding ribosomal proteins and 5S rRNA, and tRNAs, respectively. Notably, RNA polymerase I-dependent transcription represents the rate-limiting step in ribosome production [235].

Key regulators of ribosome biogenesis are the mTORC1 complex and the c-myc proto-oncogene. Both entities facilitate the expression of 47S pre-RNA by activating transcription factors for ribosomal DNA (selective factor 1) and nucleolar transcription factor (upstream binding), thereby stabilizing the initiation complex through promoter binding [214,231]. Additionally, c-myc stimulates RNA polymerase III activity via transcription factor-IIIB [231]. At promoter regions, c-myc induces histone acetylation (H3 and H4), leading to chromatin decondensation and facilitating rRNA gene transcription [236,237]. Furthermore, c-myc aids in the activation of both small and large ribosomal subunit proteins. The expression of c-myc is dependent on the phosphorylation/dephosphorylation dynamics of the Wnt/GSK-3β/β-catenin signaling pathway [226], which may impact the efficiency of ribosome biogenesis, although further experimental validation is required.

## 10. Regulation of translation by Amino acid deprivation and mTOR

During dietary restriction (DR), protein synthesis is decreased, likely due to reduced activation of the mechanistic target of rapamycin (mTOR) kinase [238,239]. A life-long reduction in protein translation, however, slows down aging, prolongs lifespan, and ameliorates cellular senescence and several age-related diseases [240-244]. The crucial response of mRNA translation to stress and longevity is established by observing that knockdown of components of the mTOR pathway or translation initiation factors directly increases longevity in various species, as does reduced S6K activity [245-249]. For example, in *S. cerevisiae*, inhibition or deletion of ribosomal protein expression increases replicative lifespan [250]. In Drosophila, the d4EBP-1 protein is vital for longevity extension under dietary restriction (DR), and its transcription regulation is under the control of the FOXO transcription factor [239,251]. FOXO transcription factor is a conserved longevity regulator downstream of insulin signaling. Insulin signaling can control TOR activity through phosphorylation of S6K and altering the activity of the upstream TSC1/TSC2 regulatory complex. Reduced Insulin/IGF-1 signaling in long-lived Ames dwarf mice and Snell mice leads to reduced protein synthesis [252,253]. Furthermore, in *C. elegans*, deletion or inhibition of several genes of translation regulators, mainly worm homologs of eIF4E (*ife-2*), eIF4G (*ifg-1*), eIF2B (*iftb-1*), and a number of ribosomal proteins, translation initiation factors increase lifespan [245,254]. Inhibition of *C. elegans*, *ifg-1*, results in selective expression of stress response genes and is also down-regulated upon starvation and reduced in long-lived dauers [240,255]. DR is also capable of activating proteins that respond to stress, which may also lead to an improvement in lifespan. Ribosomal protein and S6K knockdowns are independent of DAF-16, while the translation initiation factor may be dependent on DAF-16 to improve longevity, suggesting a complex relationship between the insulin signaling pathway and translation regulation.

To detect and respond to nutrient variations, all species continuously track their immediate environment, and a wide variety of adaptive mechanisms have evolved. How cells respond to nutritional availability and deficiencies by altering genomic information flow, transcriptional and translational levels, which are linked to global protein synthesis inhibition [256,257]. Deprivation of certain nutrients may impose more stringent restrictions on mRNA translation than others to support a particular metabolic function [258]. Amino acids play a prominent role among the key nutrients in controlling the mTORC1 pathway. The amino acids, leucine, and arginine in particular, have been reported to be completely necessary for mTORC1 activation in mammalian cells [259]. Glucose [260], glycine [164], glutamine [261], leucine [262] and serine [262] have been documented to be differently dependent on cells during division [263]. A restriction of amino acids, however, can also contribute to the activation of stress response pathways and to an improvement in model organism longevity [263,265]. In rodents, it is well known that a diet containing reduced small quantities of methionine substantially increases 45% of the safe lifespan relative to control rats [266,267]. The comprehensive mechanism related to the regulation of translation by mTOR and amino acid deprivation has been illustrated in **Fig 7.**

mTORC1 senses nutrient levels through a sophisticated system [268,269]. When cellular amino acid levels are ample, mTORC1 is activated. The discovery of the Rag GTPases has solved the mystery of how amino acids communicate their availability to mTORC1, as they are an essential component of the nutrient sensing machinery [270,271]. Rag GTPases directly bind to amino acids or their derivatives and relay that signal to mTORC1. Certain amino acids appear to be more essential for mTORC1 signaling than others; for example, leucine starvation inhibits mTORC1 and Sestrin2 [272,273]. Mechanistically, monomeric Sestrin2 binds to and antagonizes GATOR2 after leucine starvation, resulting in mTORC1 inhibition. Sestrin2 directly binds to leucine, leading to its dissociation from GATOR2 [274]. Similarly, the arginine sensor CASTOR1 directly binds to cytosolic arginine, inhibiting mTORC1 by its interaction with GATOR2 [275]. Interestingly, the sensor SLC38A9 allows arginine to interact with mTORC1 by transporting arginine-gated lysosomal amino acids, thereby enabling mTORC1 to sense both cytosolic and lysosomal arginine levels [276-278]. The amino acid transporters SLC1A5 and SLC7A5 in the plasma membrane supply cytosolic amino acids that activate mTORC1 signaling. In the absence of cytosolic amino acids, mTORC1 signaling is inhibited, leading to the activation of autophagy processes and protein degradation to replenish lysosomal amino acid pools [279,280]. Currently, it is unclear how mTORC1 activation is affected by other amino acids or the role of general amino acid sensors such as GCN2 and ATF4 in acute mTORC1 signaling cascades. While prolonged amino acid deprivation is thought to activate mTORC1 via GCN2 and ATF4, controlling the transcriptional upregulation by Sestrins, it is uncertain whether mTORC1 is regulated by GCN2 and ATF4 in transiently starved cells.

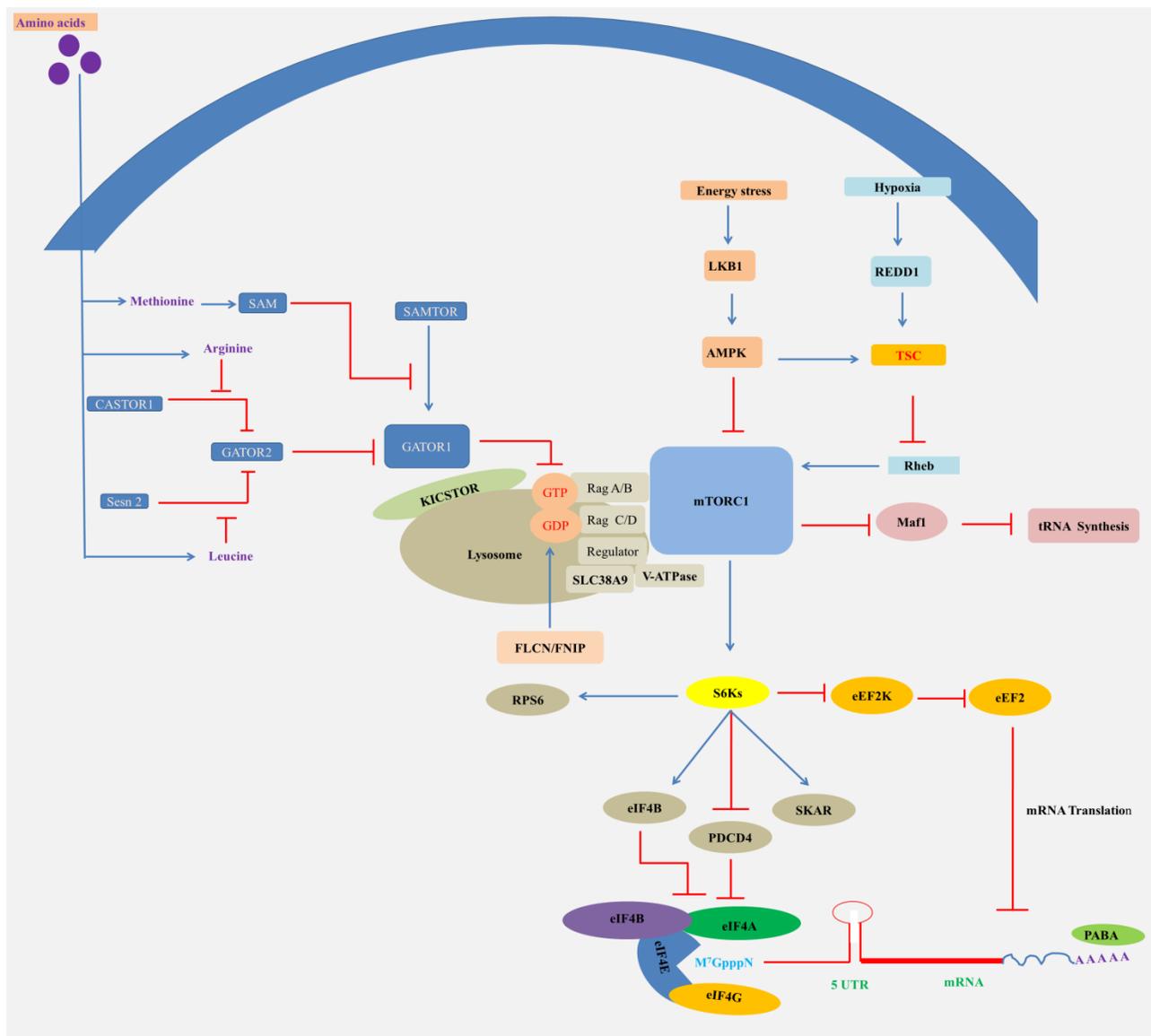

**Figure 7: Schematic representation of mTOR signaling to the translational machinery:** Rag GTPases, activated by amino acids, recruit mTORC1 to the surface of the lysosome, and the small GTPase Rheb activates mTORC1 in its GTP-bound state. The availability of amino acids controls the nucleotide state of the Rags; this process depends on the interplay between Ragulator and GATOR1. Ragulator serves as a lysosomal scaffold for RagA/B, and GATOR1 acts as a GTPase-activating protein (GAP) for RagA/B. GATOR1 is a critical negative regulator of the mTORC1 pathway. The GATOR2 complex acts in parallel to GATOR1 and is a key positive regulator of the mTORC1 pathway. The amino acid sensors Sestrin2 and SLC38A9 sense cytosolic leucine and putative lysosomal arginine, respectively, for the mTORC1 pathway. SAMTOR and CASTOR2 sense methionine and arginine, respectively, and starvation of these amino acids inhibits the mTORC1 pathway. In the absence of leucine, Sestrin2 interacts with GATOR2 and inhibits mTORC1 signaling, while SLC38A9 forms a supercomplex with Ragulator and is necessary for transmitting arginine, but not leucine, sufficiency to mTORC1. Following activation of the Ras/ERK pathway, S6K phosphorylates rpS6, eIF4B, PDCD4, and eEF2K, which are important regulators of translation. Low oxygen and energy conditions also diminish protein synthesis. Hypoxia requires the tuberous sclerosis complex (TSC) to downregulate S6K activity. In addition, translation initiation is inhibited under hypoxic conditions by the eIF2α-phosphorylating kinase PERK. Low cellular energy levels activate AMPK, which inhibits mTOR by stimulating TSC2 function.

### Acknowledgements

This work was supported by the grant from the NIH (R01AG062575-03S1). In addition, research reported in this publication was supported by an Institutional Development Award (IDeA) from the National Institute of General Medical Sciences of the National Institutes of Health under grant numbers P20GM103423 and P20GM104318. Some strains were provided by the CGC, which is funded by NIH Office of Research Infrastructure Programs (P40 OD010440).

*Conflict of Interest*: None of the authors mentioned above have conflict of interest.